\documentclass[english,aps,preprint] {revtex4}

\usepackage{graphicx}
\usepackage{amssymb}
\usepackage{amsmath}
\usepackage{setspace}
\usepackage{babel}
\usepackage{epstopdf}
\usepackage{color}
\newcommand{\Vds}{$V_{\rm{DS}}$}
\newcommand{\Ids}{$I_{\rm{DS}}$}
\newcommand{\VGone}{$V_{\rm{G1}}$}
\newcommand{\VGtwo}{$V_{\rm{G2}}$}
\newcommand{\CGone}{$C_{\rm{G1}}$}
\newcommand{\CGtwo}{$C_{\rm{G2}}$}
\newcommand{\Cs}{$C_{\rm{S}}$}
\newcommand{\Cd}{$C_{\rm{D}}$}
\newcommand{\Conetwo}{$C_{\rm{12}}$}
\newcommand{\Gs}{$\Gamma_S$}
\newcommand{\Gd}{$\Gamma_D$}
\newcommand{\Gonetwo}{$\Gamma_{12}$}

\begin{document}

\title{Pauli Blockade in a Few-Hole PMOS Double Quantum Dot limited by Spin-Orbit Interaction}

\author{H. Bohuslavskyi}
\affiliation{Universit\'e Grenoble Alpes,  F-38000 Grenoble, France}
\affiliation{CEA, LETI MINATEC campus, F-38000 Grenoble, France}
\affiliation{CEA, INAC-PHELIQS F-38000 Grenoble, France}
\author{D. Kotekar-Patil}
\affiliation{Universit\'e Grenoble Alpes,  F-38000 Grenoble, France}
\affiliation{CEA, INAC-PHELIQS F-38000 Grenoble, France}
\author{R. Maurand}
\affiliation{Universit\'e Grenoble Alpes,  F-38000 Grenoble, France}
\affiliation{CEA, INAC-PHELIQS F-38000 Grenoble, France}
\author{A. Corna}
\affiliation{Universit\'e Grenoble Alpes,  F-38000 Grenoble, France}
\affiliation{CEA, INAC-PHELIQS F-38000 Grenoble, France}
\author{S. Barraud}
\affiliation{Universit\'e Grenoble Alpes,  F-38000 Grenoble, France}
\affiliation{CEA, LETI MINATEC campus, F-38000 Grenoble, France}
\author{L. Bourdet}
\affiliation{Universit\'e Grenoble Alpes,  F-38000 Grenoble, France}
\affiliation{CEA, INAC-MEM F-38000 Grenoble, France}
\author{L. Hutin}
\affiliation{Universit\'e Grenoble Alpes,  F-38000 Grenoble, France}
\affiliation{CEA, LETI MINATEC campus, F-38000 Grenoble, France}
\author{Y.-M. Niquet}
\affiliation{Universit\'e Grenoble Alpes,  F-38000 Grenoble, France}
\affiliation{CEA, INAC-MEM F-38000 Grenoble, France}
\author{X. Jehl}
\affiliation{Universit\'e Grenoble Alpes,  F-38000 Grenoble, France}
\affiliation{CEA, INAC-PHELIQS F-38000 Grenoble, France}
\author{S. De Franceschi}
\affiliation{Universit\'e Grenoble Alpes,  F-38000 Grenoble, France}
\affiliation{CEA, INAC-PHELIQS F-38000 Grenoble, France}
\author{M. Vinet}
\affiliation{Universit\'e Grenoble Alpes,  F-38000 Grenoble, France}
\affiliation{CEA, LETI MINATEC campus, F-38000 Grenoble, France}
\author{M. Sanquer}
\affiliation{Universit\'e Grenoble Alpes,  F-38000 Grenoble, France}
\affiliation{CEA, INAC-PHELIQS F-38000 Grenoble, France}

\begin{abstract}
We report on hole compact double quantum dots  fabricated using conventional CMOS technology. We provide evidence of  Pauli spin blockade in the few hole regime which is relevant to spin qubit implementations.
 A current dip is observed around zero magnetic field, in agreement with the expected behavior  for the case of  strong spin-orbit. We deduce an intradot spin relaxation rate $\approx$120\,kHz for the first holes, an important step towards a robust hole spin-orbit qubit.

\end{abstract}
\maketitle

Since the proposal of Loss and DiVincenzo in 1998~\cite{Loss1998} to make quantum bits based on spins confined in semiconductor quantum dots, substantial progress has been made. First in III-V materials, where the maturity of growth techniques has allowed the emergence of top-down qubits based on the confinement of a two-dimensional electron gas in GaAs/AlGaAs hetero-structures~\cite{Hanson2007}, but also of bottom-up qubits made from nanowires (InAs or InSb)~\cite{Nadj-Perge2010a,Berg2013}. In all III-V qubits, the dephasing time is limited by the interaction of the electron spin with the nuclear spins present in the host material~\cite{Bluhm2011,Lange2010}. In contrast silicon, which presents a low natural abundance of nuclear spins and can even be isotopically purified, can be used to make electron spin qubits with extremely long dephasing time~\cite{Maune2012, Wu2014, Kawakami2014, Veldhorst2014}. An all-electrical control of single dot spin qubit by a single gate voltage microwave signal without the need of local magnetic field gradient \cite{Pioro-Ladriere2008} would be a clear asset for future developments.  Fast and local electrical manipulation using spin-orbit interactions has already been demonstrated in III-V materials.~\cite{Nadj-Perge2010a,Berg2013}. Thus focusing on holes in silicon appears as an appealing strategy since valence-band holes present a limited hyperfine interaction\cite{Testelin2009} together with a strong spin-orbit interaction (SOI) due to their p-orbital nature. Recent experiments~\cite{Li2015,Voisin2016} have indeed revealed SOI-related spin properties and a hole spin qubit has even been demonstrated~\cite{Maurand2016}. Here, we report on the implementation of a silicon hole double quantum dot (DQD)  based on the technology decribed in refs.~\onlinecite{Barraud2012a,Barraud2016}. The device is tunable in the few hole regime in which we investigate Pauli spin blockade (PSB), the key ingredient for spin qubits initialization and readout in several qubit implementations~\cite{Hanson2007}.  More specifically, we focus on the magnetic field evolution of the leakage current through the device in the PSB regime. It reveals a dip around zero magnetic field linked to spin-orbit mixing~\cite{Danon2009}.
The spin relaxation rates determined from the PSB are comparable with the values extracted for electrons in InAs nanowire double quantum dots~\cite{Nadj-Perge2010} and are compatible with the operation of a hole spin-orbit qubit in silicon~\cite{Maurand2016}.

\begin{figure}
 \begin{center}
\includegraphics[width=\columnwidth]{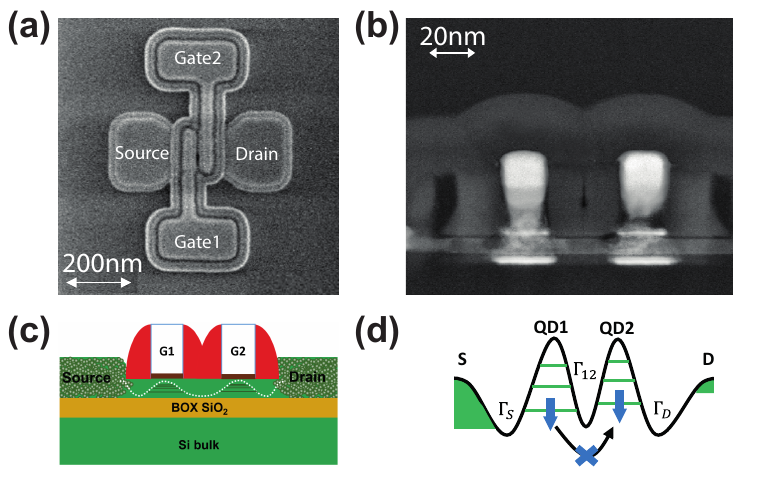}
 \end{center}
\caption{(a) Top view scanning electron micrograph of a typical DQD device after spacer etching, featuring 30\,nm-long gates separated by 35\,nm. 
(b) Transmission electron micrograph along the source-drain axis. 
(c) Schematics of the DQD made by hole accumulation below G1 and G2. 
(d) Schematic Pauli spin blockade for the (1h,1h)$\rightarrow$(0,2h) transition at reverse bias (\Vds $\le$0, analogue to the (3h,3h)$\rightarrow$(2h,4h) transition in the inset of Fig.~\ref{fig:fig4}c). The black line is the top of the valence band. The green regions indicate the hole reservoirs in the source and drain. }
\label{fig:fig1}
\end{figure}

Our devices are  nanowire field-effect transistors fabricated in a 300\,mm CMOS facility on silicon-on-insulator  wafers with 145 nm-thick buried oxide. The 11\,nm-thick Si channel is doped with   $ \simeq 4\times 10^{24}$ Boron.m$^{-3}$. Nanowire width down to 18 nm are achieved after patterning.
Two gates (G1 and G2) in series are patterned by electron-beam lithography and are isolated from the channel by 2.5\,nm of SiO$_2$ and 1.9\,nm of HfO$_2$. Silicon nitride spacers are then deposited and etched on the sidewalls of the gates (see figure \ref{fig:fig1}a,b). The spacers effectively protect the inter-gate spacing from the silicidation and dopant implantation used to reduce the access resistances~\cite{Barraud2016}.
The resulting structure, sketched in figure~\ref{fig:fig1}c, yields a compact DQD with optimal gate control. At low temperature, two quantum dots, QD1 and QD2, are formed by accumulation below G1 and G2 respectively (see figure~\ref{fig:fig1}c-d). The same process has been used to produce n-type DQD\cite{Kotekar2016}.

\begin{figure}
 \begin{center}
\includegraphics[width=10cm]{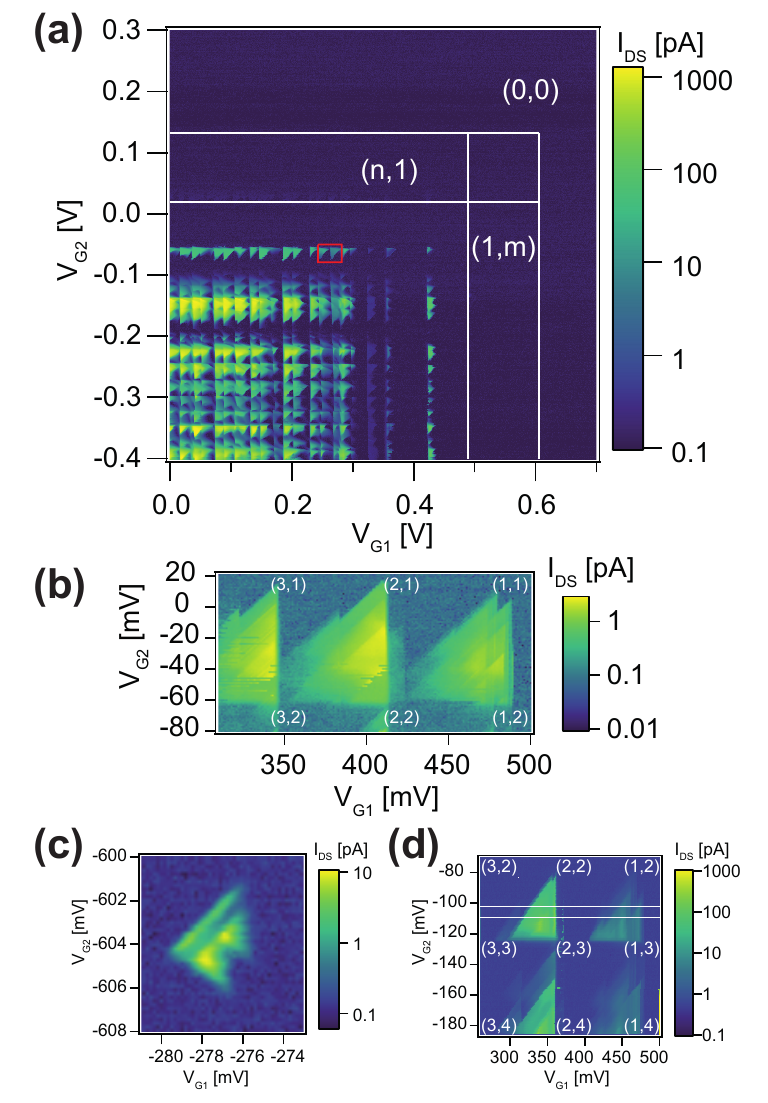}
 \end{center}
\caption{(a) \Ids{} versus \VGone{} and \VGtwo{} measured with \Vds{}=20\,mV at T=60\,mK. White lines indicate the position of four lines of current detected at larger bias (see b). The absolute hole occupation numbers are indicated for the few hole region. The red square indicates the region studied in fig.~\ref{fig:fig3}a). (b) \Ids{} versus \VGone{} and \VGtwo{}  measured with \Vds{}=-100\,mV at T=60\,mK in a region where no current is detected at \Vds{}=20\,mV ((n,1)$\rightarrow$(n,2) transition line).  (c)\Ids{} versus \VGone{} and \VGtwo{} measured with \Vds{}=-3\,mV at T=60\,mK in the many hole regime. (d) \Ids{} versus \VGone{} and \VGtwo{} measured with \Vds{}=-70\,mV at T=60\,mK in the region where PSB has been studied (see Fig. \ref{fig:fig4}). The absolute hole occupation numbers are indicated.}
\label{fig:fig2}
\end{figure}

Electrical  characterization was performed from T=300\,K down to very low temperature by recording the drain-source current \Ids{} as a function of the two gate voltages \VGone{} and \VGtwo{} (stability diagram) at various drain source voltages \Vds. In the experimental setup the source was grounded.
The stability diagram shown in fig.~\ref{fig:fig2}a reveals overlapped bias triangles~\cite{Wiel2002} with vertical and horizontal edges. This is a characteristic of an excellent electrostatic control of each dot by one gate. The gate capacitances associated  to G1 and G2, \CGone{} and \CGtwo{} are therefore the dominant capacitances and the lever arm parameters  $\frac{C_{G1}}{C_{\Sigma1}}$  and $\frac{C_{G2}}{C_{\Sigma2}}$ are close to 1 ($C_{\Sigma1}$=\CGone+\Cs+\Conetwo{} and $C_{\Sigma2}$=\CGtwo+\Cd+\Conetwo{} are the total capacitances for QD1 and QD2, \Conetwo{} being the capacitance between QD1 and QD2 and \Cs{} (\Cd{}) being the source to QD1 (drain to QD2) capacitance).

In order to precisely know $(n,m)$  
-the charge state with $n(m)$ excess holes in QD1(QD2)- a large \Vds{} has been applied.
Even if transitions $(1,m)\rightarrow(0,m+1)$ and $(n,1)\rightarrow(n+1,0)$ have not been detected at \Vds{}=20\,mV, they appear above  $\vert$ \Vds{} $\vert$  $\approx$ 100\,mV  thanks to the enhanced tunneling through the barriers under the spacers: the latter are markedly tilted at high \Vds{} so that  the tunnel transparencies \Gs{},  \Gd{}  increase significantly.  In  figure~\ref{fig:fig2}b, the drain current recorded at \Vds{}=-100\,mV  is shown in a region where no current is detected at \Vds{}=20\,mV. This row of triangles correspond to the second (1$\rightarrow$2) transition in  dot 2. 
The conducting parts of the triangles are replicated as a result of the ionization of dopants near the channel at large bias\cite{Golovach2011}.

Interestingly, the charging energies are significantly larger in the few holes (up to 70 meV) than in the many holes regime ($\simeq$ 20 meV). We have, therefore, performed tight-binding calculations \cite{Niquet2009} in a realistic geometry in order to understand the nature of the very first low-lying hole states. The first few holes do not localize in edge states as in Ref. \cite{edges_NL2014}  because the channel is doped with Boron atoms and the back gate is grounded, therefore the hole are not pulled in the upper corners. They might rather be bound to clusters of two or more nearby Boron impurities  which exist in the doped channel. Assuming a random distribution of Boron atoms, there is indeed $>50$\% (resp. $>95$\%) chance of having at least two impurities closer than $d=1.5$ nm (resp. $d=2.5$ nm) under the gate. Configuration interaction calculations show that such clusters show larger binding and charging energies $E_c$ than single impurities ($E_c\sim 75$ meV at $d=1.5$ nm and $E_c<60$ meV when $d>2.5$ nm).  The charging energy decreases once the deepest clusters are filled and the confinement gets dominated by the structure and gate fields. Despite doping, the SOI is mostly mediated by the silicon matrix as the probability that the holes sit on the Boron atoms is always small.
 
\begin{figure}
 \begin{center}
\includegraphics[width=\columnwidth]{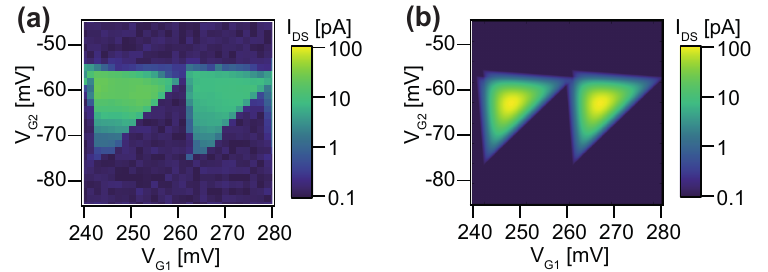}
 \end{center}
\caption{(a) \Ids{} versus \VGone{} and \VGtwo{} measured with \Vds{}=20\,mV at T=60\,mK (region highlighted in red in fig.~\ref{fig:fig2}a) (b)Electrostatic simulations with the parameters given in table~\ref{table}.}
\label{fig:fig3}
\end{figure}

Once the first holes are added in the channel the DQD is defined by the geometry of the sample. 
We have simulated the stability diagram in the $(n,m)\ge(5,3)$ regime with the orthodox Coulomb blockade theory. We solved the master equation for transport \cite{Pierre2009} with the parameters given in table~\ref{table}. In addition to the capacitances defined above, we set the electronic temperature $T_e$, as well as the tunneling rates \Gs{}, \Gd{} and \Gonetwo{}  associated to \Cs{}, \Cd{} and \Conetwo respectively. 
The  simulation, shown in fig.~\ref{fig:fig3}b, reproduces the shape of the measured bias triangles.

\begin{table}
\begin{tabular}{|c|c|}
\hline $T_e$ & 150\,mK \\ 
\hline \CGone=\CGtwo & 7.6\,aF \\ 
\hline \Cs=\Cd & 0.15\,aF  \\
\hline \Conetwo & 0.65\,aF \\ 
\hline \Gs=\Gd=\Gonetwo & $10^{-4}$\,$\frac{e^2}{h}$ \\  
\hline
\end{tabular} 
\caption{Numerical values used in the simulation of fig. \ref{fig:fig3}b.}
\label{table}
\end{table}

The value for \Conetwo{} is deduced from the gate voltage separation $\Delta V_{\rm G}$ between the triple points~\cite{Wiel2002} observed at small \Vds (see fig. \ref{fig:fig2}c):
$\Delta V_{\rm G}= e\frac{C_{\rm 12}}{C_{\rm G1}C_{\rm G2}} \simeq 1.8$\,mV.
The values of \CGone{} and \CGtwo{} used in this simulation are in good agreement with a planar model for the gate capacitance of our DQD: $C_{\rm G1(2)}={\frac{\epsilon_0\epsilon_{SiO_{2}}A_{1(2)}}{EOT}}\approx$ 11\,aF, where $A_{1(2)}$ is  the  channel area covered by G1 (G2), and $EOT \simeq 2.9$ is the equivalent oxide thickness  nm~\cite{Hofheinz2006}. 

We now turn to the investigation of PSB. Spin blockade in a DQD arises when the current involves a transport cycle equivalent to $(0,1)\to(1,1)\to(0,2)\to(0,1)$~\cite{Ono2002}. Since the $(0,2)$ ground state is a spin singlet, the cycle stops as soon as the DQD enters in a (1,1) triplet state. The remaining leakage current results from spin relaxation or spin-orbit mixing mechanisms. Depending on the relevant mechanism, the leakage current will behave differently as a function of the magnetic field and detuning\cite{Danon2009,Nadj-Perge2010}.

\begin{figure}
 \begin{center}
\includegraphics[width=\columnwidth]{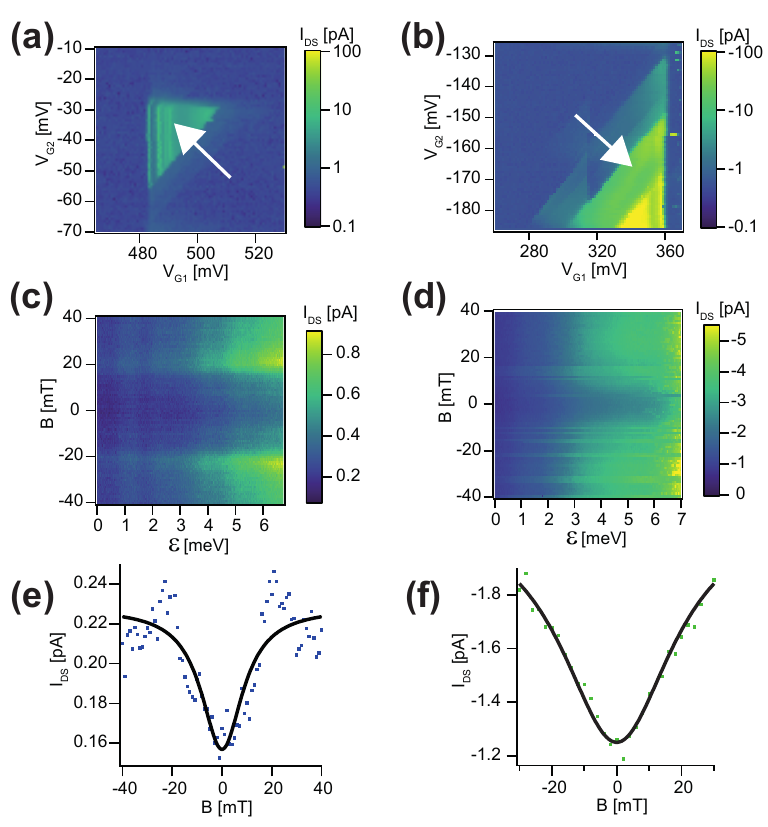}
 \end{center}
\caption{Current in the PSB regime as a function of detuning and out-of-plane magnetic field $B$ at $T$=60\,mK. (a) Current versus \VGone{} and \VGtwo{} at \Vds=70\,mV and $B$=0. (b) Same as in (a) except for \Vds=-70\,mV. (c) Current versus detuning energy $\epsilon$ and magnetic field for the (1,3)$\rightarrow$(2,2) transition (white arrow in a)).(d) Same as in (c) but for the (3,3)$\rightarrow$(2,4) transition (white arrow in b)). (e) and (f) are cuts of the (1,3)$\rightarrow$(2,2) and (3,3)$\rightarrow$(2,4) transitions at $\epsilon$=0. The curves are fitted (black lines) assuming that PSB is spin-orbit mediated~\cite{Danon2009}.}
\label{fig:fig4}
\end{figure}

  Figs.~\ref{fig:fig4}a and \ref{fig:fig4}b present current triangles in which  PSB is evidenced at $T=60$\,mK thanks to the magnetic field dependence of the drain-source current.  As expected, a reduced  current is  detected at the base of the bias triangles\cite{Hanson2007} corresponding to the $(1,3)\to(2,2)$ transition in Fig.~\ref{fig:fig4}a and to the $(3,3)\to(2,4)$ transition in Fig.~\ref{fig:fig4}b, respectively.  Figs.~\ref{fig:fig4}c and \ref{fig:fig4}d display the leakage current as a function of the out-of-plane magnetic field, $B$, and of the detuning axis in the PSB regime of Fig.~\ref{fig:fig4}a and Fig.~\ref{fig:fig4}b, respectively (detuning axis are indicated by  white arrows in Figs.~\ref{fig:fig4}a and \ref{fig:fig4}d). The leakage current decreases around $B=0$  in both cases. The  current does not depend on magnetic field for the reverse polarity  ($V_{\rm DS}<0$) as well as for the two other triangles  shown in Fig. \ref{fig:fig2}d, i.e. for $(3,2) \leftrightarrow (2,3)$ and $(2,3) \leftrightarrow (1,4)$ transitions. Note here that the $(1,1)\to(0,2)$ transition - at which PSB is also expected- was not caught even at large bias. 
  
  A cut at zero detuning taken in Fig.~\ref{fig:fig4}c (in Fig.~\ref{fig:fig4}d) is shown in figure Fig.~\ref{fig:fig4}e (in Fig.~\ref{fig:fig4}f). It reveals a current dip that can be  fitted to a Lorentzian function, in line with a model assuming  strong SOI\cite{Danon2009}:
\begin{equation}
I = I_{\rm max}(1-{\frac{8}{9}}{\frac{B_C^2}{B_C^2+B^2}})+I_0
\label{eq:crossover}
\end{equation}
with $I_{\rm max} =4e\Gamma_{\rm rel}$ the dip height, where $\Gamma_{\rm rel}$ is the spin relaxation rate among the (1,1) states, $B_{\rm C}$ is the dip width and $ I_{0}$ is a  B-independent background  current~\cite{Li2015}($0.15$ pA for the (1,3)$\rightarrow$(2,2) transition and $1.3$ pA for the (3,3)$\rightarrow$(2,4) transition). $B_{\rm C}$ accounts for the cross-over between leakage currents resulting from spin relaxation at small field and spin-orbit mixing at higher field. The rate $\Gamma_{\rm SO}$ of spin-orbit mixing between (1,1) states and the (0,2) singlet can be estimated with:
\begin{equation}
g\mu_{\rm B} B_{\rm C} \simeq h \sqrt{\Gamma_{\rm rel}\times\Gamma_{\rm SO} }
\label{eq:ct}
\end{equation}

Contrarily to refs.~\onlinecite{Yamahata2012,Li2015}, we always see a dip of current at low magnetic field that we attribute to the dominance of spin-orbit mixing over hyperfine~\cite{Nadj-Perge2010} or spin-flip cotunneling mechanisms~\cite{Yamahata2012,Morello_2011SB}.
We also observe two current peaks at $B=\pm20$\,mT and $T= 60$\,mK (see Figs.~\ref{fig:fig4}c and \ref{fig:fig4}e). Peaks of current at finite magnetic field, whose origin remains unclear, are also reported in refs.~\onlinecite{Pfund2007,Yamahata2012,S.2013}. The dip observed at zero magnetic field extends in detuning up to several meV, which indicates that the $(0,2)$ singlet-triplet splitting in our QDs -as other orbital splittings - is large. As a result PSB can be seen even at $T$=4.2\,K (not shown).

 $\Gamma_{rel}$ and $\Gamma_{SO}$ can be estimated from the above experiments. For two QDs in series, $\Gamma_{SO}$ mainly depends on the interdot coupling. The hole $g$-factor was found to be anisotropic in similar nanowire transistors~\cite{Voisin2016}, with $g=$1.5-2.6. Eq. (\ref{eq:ct}) then yields $\Gamma_{SO}$=1.4-4.3\,meV for the (1,3)$\rightarrow$(2,2) transition and $\Gamma_{SO}$=0.6-1.8\,meV  for the (3,3)$\rightarrow$(2,4) transition. The spin relaxation is dominated by the spin-orbit coupling in our DQD rather than by hyperfine effects. Indeed we estimate a fluctuating Overhauser field $B_{nuc}\approx$20\,$\rm{\mu}$T, which is much smaller than the current dip width of $10$-$20$\,mT~\cite{Yamahata2012,Voisin2016}. $\Gamma_{SO}$ is larger than in previous reports~\cite{Nadj-Perge2010,Li2015} while the critical field $B_C$ is comparable to that of refs.~\onlinecite{Nadj-Perge2010,Yamahata2012} (for electrons) and smaller than in refs.~\onlinecite{S.2013,Li2015} (for holes). This can be attributed to the small value of $\Gamma_{rel}$. A large $\Gamma_{SO}$ can limit qubit readout fidelity through unwanted transitions from  (1,1) triplet to  (0,2) singlet~\cite{Nadj-Perge2012} and it would be favorable to reduce the interdot coupling.

$\Gamma_{rel} = 120$ kHz (resp.  $ 2.0$ MHz) for the (1,3)$\rightarrow$(2,2) (resp. (3,3)$\rightarrow$(2,4))  transition  is smaller than in previous reports~\cite{Nadj-Perge2010,Yamahata2012,S.2013,Li2015} where it ranges between 0.8~\cite{Yamahata2012} and 6\,MHz \cite{Nadj-Perge2012} (3\,MHz in ref.~\onlinecite{Li2015}, \Ids=6\,pA for $B\ge B_C$ in ref.~\onlinecite{S.2013}). $\Gamma_{rel}$, which limits the inelastic relaxation time $T_1$, should be primarily minimized for hole spin-orbit qubits. Contrarily to $\Gamma_{SO}$, $\Gamma_{rel}$ cannot be adjusted by changing the interdot coupling and its optimization is material and process dependent.

To conclude, the few-hole regime has been reached in a silicon CMOS DQD and Pauli spin blockade has been observed at different charge transitions.We found that this blockade is dominated by the SOI. By analyzing the magnetic field evolution of the leakage current in the blockade regime we deduced a small intradot spin relaxation rate ($\approx$120\,kHz for the first holes), an important step towards a robust hole spin-orbit qubit.

\section*{Acknowledgments}
The authors thank C. Guedj and G. Audoit for extensive TEM analysis. This work is supported by the EU through the FP7 ICT  SiSPIN (323841), SiAM (610637) and H2020 ICT25 MOSQUITO (688539) collaborative projects. Part of the calculations were run on TGCC/Curie thanks to a GENCI allocation.
\bibliography{biblio_holeDQDv4}

\begin{thebibliography}{32}
\expandafter\ifx\csname natexlab\endcsname\relax\def\natexlab#1{#1}\fi
\expandafter\ifx\csname bibnamefont\endcsname\relax
  \def\bibnamefont#1{#1}\fi
\expandafter\ifx\csname bibfnamefont\endcsname\relax
  \def\bibfnamefont#1{#1}\fi
\expandafter\ifx\csname citenamefont\endcsname\relax
  \def\citenamefont#1{#1}\fi
\expandafter\ifx\csname url\endcsname\relax
  \def\url#1{\texttt{#1}}\fi
\expandafter\ifx\csname urlprefix\endcsname\relax\def\urlprefix{URL }\fi
\providecommand{\bibinfo}[2]{#2}
\providecommand{\eprint}[2][]{\url{#2}}

\bibitem[{\citenamefont{Loss and DiVincenzo}(1998)}]{Loss1998}
\bibinfo{author}{\bibfnamefont{D.}~\bibnamefont{Loss}} \bibnamefont{and}
  \bibinfo{author}{\bibfnamefont{D.~P.} \bibnamefont{DiVincenzo}},
  \bibinfo{journal}{Physical Review A} \textbf{\bibinfo{volume}{57}},
  \bibinfo{pages}{120} (\bibinfo{year}{1998}).

\bibitem[{\citenamefont{Hanson et~al.}(2007)\citenamefont{Hanson, Kouwenhoven,
  Petta, Tarucha, and Vandersypen}}]{Hanson2007}
\bibinfo{author}{\bibfnamefont{R.}~\bibnamefont{Hanson}},
  \bibinfo{author}{\bibfnamefont{L.~P.} \bibnamefont{Kouwenhoven}},
  \bibinfo{author}{\bibfnamefont{J.~R.} \bibnamefont{Petta}},
  \bibinfo{author}{\bibfnamefont{S.}~\bibnamefont{Tarucha}}, \bibnamefont{and}
  \bibinfo{author}{\bibfnamefont{L.~M.~K.} \bibnamefont{Vandersypen}},
  \bibinfo{journal}{Rev. Mod. Phys.} \textbf{\bibinfo{volume}{79}},
  \bibinfo{pages}{1217} (\bibinfo{year}{2007}),
  \urlprefix\url{http://link.aps.org/doi/10.1103/RevModPhys.79.1217}.

\bibitem[{\citenamefont{Nadj-Perge
  et~al.}(2010{\natexlab{a}})\citenamefont{Nadj-Perge, Frolov, Bakkers, and
  Kouwenhoven}}]{Nadj-Perge2010a}
\bibinfo{author}{\bibfnamefont{S.}~\bibnamefont{Nadj-Perge}},
  \bibinfo{author}{\bibfnamefont{S.~M.} \bibnamefont{Frolov}},
  \bibinfo{author}{\bibfnamefont{E.~P. A.~M.} \bibnamefont{Bakkers}},
  \bibnamefont{and} \bibinfo{author}{\bibfnamefont{L.~P.}
  \bibnamefont{Kouwenhoven}}, \bibinfo{journal}{Nature}
  \textbf{\bibinfo{volume}{468}}, \bibinfo{pages}{1084}
  (\bibinfo{year}{2010}{\natexlab{a}}), ISSN \bibinfo{issn}{0028-0836},
  \urlprefix\url{http://dx.doi.org/10.1038/nature09682}.

\bibitem[{\citenamefont{van~den Berg et~al.}(2013)\citenamefont{van~den Berg,
  Nadj-Perge, Pribiag, Plissard, Bakkers, Frolov, and Kouwenhoven}}]{Berg2013}
\bibinfo{author}{\bibfnamefont{J.~W.~G.} \bibnamefont{van~den Berg}},
  \bibinfo{author}{\bibfnamefont{S.}~\bibnamefont{Nadj-Perge}},
  \bibinfo{author}{\bibfnamefont{V.~S.} \bibnamefont{Pribiag}},
  \bibinfo{author}{\bibfnamefont{S.~R.} \bibnamefont{Plissard}},
  \bibinfo{author}{\bibfnamefont{E.~P. A.~M.} \bibnamefont{Bakkers}},
  \bibinfo{author}{\bibfnamefont{S.~M.} \bibnamefont{Frolov}},
  \bibnamefont{and} \bibinfo{author}{\bibfnamefont{L.~P.}
  \bibnamefont{Kouwenhoven}}, \bibinfo{journal}{Phys. Rev. Lett.}
  \textbf{\bibinfo{volume}{110}}, \bibinfo{pages}{066806}
  (\bibinfo{year}{2013}),
  \urlprefix\url{http://link.aps.org/doi/10.1103/PhysRevLett.110.066806}.

\bibitem[{\citenamefont{Bluhm et~al.}(2011)\citenamefont{Bluhm, Foletti, Neder,
  Rudner, Mahalu, Umansky, and Yacoby}}]{Bluhm2011}
\bibinfo{author}{\bibfnamefont{H.}~\bibnamefont{Bluhm}},
  \bibinfo{author}{\bibfnamefont{S.}~\bibnamefont{Foletti}},
  \bibinfo{author}{\bibfnamefont{I.}~\bibnamefont{Neder}},
  \bibinfo{author}{\bibfnamefont{M.}~\bibnamefont{Rudner}},
  \bibinfo{author}{\bibfnamefont{D.}~\bibnamefont{Mahalu}},
  \bibinfo{author}{\bibfnamefont{V.}~\bibnamefont{Umansky}}, \bibnamefont{and}
  \bibinfo{author}{\bibfnamefont{A.}~\bibnamefont{Yacoby}},
  \bibinfo{journal}{Nat Phys} \textbf{\bibinfo{volume}{7}},
  \bibinfo{pages}{109} (\bibinfo{year}{2011}), ISSN \bibinfo{issn}{1745-2473},
  \urlprefix\url{http://dx.doi.org/10.1038/nphys1856}.

\bibitem[{\citenamefont{de~Lange et~al.}(2010)\citenamefont{de~Lange, Wang,
  Rist{\`e}, Dobrovitski, and Hanson}}]{Lange2010}
\bibinfo{author}{\bibfnamefont{G.}~\bibnamefont{de~Lange}},
  \bibinfo{author}{\bibfnamefont{Z.~H.} \bibnamefont{Wang}},
  \bibinfo{author}{\bibfnamefont{D.}~\bibnamefont{Rist{\`e}}},
  \bibinfo{author}{\bibfnamefont{V.~V.} \bibnamefont{Dobrovitski}},
  \bibnamefont{and} \bibinfo{author}{\bibfnamefont{R.}~\bibnamefont{Hanson}},
  \bibinfo{journal}{Science} \textbf{\bibinfo{volume}{330}},
  \bibinfo{pages}{60} (\bibinfo{year}{2010}), ISSN \bibinfo{issn}{0036-8075},
  \urlprefix\url{http://science.sciencemag.org/content/330/6000/60}.

\bibitem[{\citenamefont{Maune et~al.}(2012)\citenamefont{Maune, Borselli,
  Huang, Ladd, Deelman, Holabird, Kiselev, Alvarado-Rodriguez, Ross, Schmitz
  et~al.}}]{Maune2012}
\bibinfo{author}{\bibfnamefont{B.~M.} \bibnamefont{Maune}},
  \bibinfo{author}{\bibfnamefont{M.~G.} \bibnamefont{Borselli}},
  \bibinfo{author}{\bibfnamefont{B.}~\bibnamefont{Huang}},
  \bibinfo{author}{\bibfnamefont{T.~D.} \bibnamefont{Ladd}},
  \bibinfo{author}{\bibfnamefont{P.~W.} \bibnamefont{Deelman}},
  \bibinfo{author}{\bibfnamefont{K.~S.} \bibnamefont{Holabird}},
  \bibinfo{author}{\bibfnamefont{A.~A.} \bibnamefont{Kiselev}},
  \bibinfo{author}{\bibfnamefont{I.}~\bibnamefont{Alvarado-Rodriguez}},
  \bibinfo{author}{\bibfnamefont{R.~S.} \bibnamefont{Ross}},
  \bibinfo{author}{\bibfnamefont{A.~E.} \bibnamefont{Schmitz}},
  \bibnamefont{et~al.}, \bibinfo{journal}{Nature}
  \textbf{\bibinfo{volume}{481}}, \bibinfo{pages}{344} (\bibinfo{year}{2012}),
  ISSN \bibinfo{issn}{0028-0836},
  \urlprefix\url{http://dx.doi.org/10.1038/nature10707}.

\bibitem[{\citenamefont{Wu et~al.}(2014)\citenamefont{Wu, Ward, Prance, Kim,
  Gamble, Mohr, Shi, Savage, Lagally, Friesen et~al.}}]{Wu2014}
\bibinfo{author}{\bibfnamefont{X.}~\bibnamefont{Wu}},
  \bibinfo{author}{\bibfnamefont{D.~R.} \bibnamefont{Ward}},
  \bibinfo{author}{\bibfnamefont{J.~R.} \bibnamefont{Prance}},
  \bibinfo{author}{\bibfnamefont{D.}~\bibnamefont{Kim}},
  \bibinfo{author}{\bibfnamefont{J.~K.} \bibnamefont{Gamble}},
  \bibinfo{author}{\bibfnamefont{R.~T.} \bibnamefont{Mohr}},
  \bibinfo{author}{\bibfnamefont{Z.}~\bibnamefont{Shi}},
  \bibinfo{author}{\bibfnamefont{D.~E.} \bibnamefont{Savage}},
  \bibinfo{author}{\bibfnamefont{M.~G.} \bibnamefont{Lagally}},
  \bibinfo{author}{\bibfnamefont{M.}~\bibnamefont{Friesen}},
  \bibnamefont{et~al.}, \bibinfo{journal}{Proceedings of the National Academy
  of Sciences} \textbf{\bibinfo{volume}{111}}, \bibinfo{pages}{11938}
  (\bibinfo{year}{2014}),
  \urlprefix\url{http://www.pnas.org/content/111/33/11938.abstract}.

\bibitem[{\citenamefont{Kawakami et~al.}(2014)\citenamefont{Kawakami, Scarlino,
  Ward, Braakman, Savage, Lagally, Friesen, Coppersmith, Eriksson, and
  Vandersypen}}]{Kawakami2014}
\bibinfo{author}{\bibfnamefont{E.}~\bibnamefont{Kawakami}},
  \bibinfo{author}{\bibfnamefont{P.}~\bibnamefont{Scarlino}},
  \bibinfo{author}{\bibfnamefont{D.~R.} \bibnamefont{Ward}},
  \bibinfo{author}{\bibfnamefont{F.~R.} \bibnamefont{Braakman}},
  \bibinfo{author}{\bibfnamefont{D.~E.} \bibnamefont{Savage}},
  \bibinfo{author}{\bibfnamefont{M.~G.} \bibnamefont{Lagally}},
  \bibinfo{author}{\bibfnamefont{M.}~\bibnamefont{Friesen}},
  \bibinfo{author}{\bibfnamefont{S.~N.} \bibnamefont{Coppersmith}},
  \bibinfo{author}{\bibfnamefont{M.~A.} \bibnamefont{Eriksson}},
  \bibnamefont{and} \bibinfo{author}{\bibfnamefont{L.~M.~K.}
  \bibnamefont{Vandersypen}}, \bibinfo{journal}{Nat Nano}
  \textbf{\bibinfo{volume}{9}}, \bibinfo{pages}{666} (\bibinfo{year}{2014}),
  ISSN \bibinfo{issn}{1748-3387},
  \urlprefix\url{http://dx.doi.org/10.1038/nnano.2014.153}.

\bibitem[{\citenamefont{Veldhorst et~al.}(2014)\citenamefont{Veldhorst, Hwang,
  Yang, de~Ronde, Dehollain, Muhonen, Hudson, Itoh, Morello, and
  Dzurak}}]{Veldhorst2014}
\bibinfo{author}{\bibfnamefont{M.}~\bibnamefont{Veldhorst}},
  \bibinfo{author}{\bibfnamefont{J.~C.~H.} \bibnamefont{Hwang}},
  \bibinfo{author}{\bibfnamefont{A.~W.} \bibnamefont{Yang},
  \bibfnamefont{C.~H.~andLeenstra}},
  \bibinfo{author}{\bibfnamefont{B.}~\bibnamefont{de~Ronde}},
  \bibinfo{author}{\bibfnamefont{J.~P.} \bibnamefont{Dehollain}},
  \bibinfo{author}{\bibfnamefont{J.~T.} \bibnamefont{Muhonen}},
  \bibinfo{author}{\bibfnamefont{F.~E.} \bibnamefont{Hudson}},
  \bibinfo{author}{\bibfnamefont{K.~M.} \bibnamefont{Itoh}},
  \bibinfo{author}{\bibfnamefont{A.}~\bibnamefont{Morello}}, \bibnamefont{and}
  \bibinfo{author}{\bibfnamefont{A.~S.} \bibnamefont{Dzurak}},
  \bibinfo{journal}{Nat Nano} \textbf{\bibinfo{volume}{9}},
  \bibinfo{pages}{981} (\bibinfo{year}{2014}), ISSN \bibinfo{issn}{1748-3387},
  \urlprefix\url{http://dx.doi.org/10.1038/nnano.2014.216}.

\bibitem[{\citenamefont{Pioro-Ladriere
  et~al.}(2008)\citenamefont{Pioro-Ladriere, Obata, Tokura, Shin, Kubo,
  Yoshida, Taniyama, and Tarucha}}]{Pioro-Ladriere2008}
\bibinfo{author}{\bibfnamefont{M.}~\bibnamefont{Pioro-Ladriere}},
  \bibinfo{author}{\bibfnamefont{T.}~\bibnamefont{Obata}},
  \bibinfo{author}{\bibfnamefont{Y.}~\bibnamefont{Tokura}},
  \bibinfo{author}{\bibfnamefont{Y.-S.} \bibnamefont{Shin}},
  \bibinfo{author}{\bibfnamefont{T.}~\bibnamefont{Kubo}},
  \bibinfo{author}{\bibfnamefont{K.}~\bibnamefont{Yoshida}},
  \bibinfo{author}{\bibfnamefont{T.}~\bibnamefont{Taniyama}}, \bibnamefont{and}
  \bibinfo{author}{\bibfnamefont{S.}~\bibnamefont{Tarucha}},
  \bibinfo{journal}{Nat Phys} \textbf{\bibinfo{volume}{4}},
  \bibinfo{pages}{776} (\bibinfo{year}{2008}), ISSN \bibinfo{issn}{1745-2473},
  \urlprefix\url{http://dx.doi.org/10.1038/nphys1053}.

\bibitem[{\citenamefont{Testelin et~al.}(2009)\citenamefont{Testelin,
  Bernardot, Eble, and Chamarro}}]{Testelin2009}
\bibinfo{author}{\bibfnamefont{C.}~\bibnamefont{Testelin}},
  \bibinfo{author}{\bibfnamefont{F.}~\bibnamefont{Bernardot}},
  \bibinfo{author}{\bibfnamefont{B.}~\bibnamefont{Eble}}, \bibnamefont{and}
  \bibinfo{author}{\bibfnamefont{M.}~\bibnamefont{Chamarro}},
  \bibinfo{journal}{Phys. Rev. B} \textbf{\bibinfo{volume}{79}},
  \bibinfo{pages}{195440} (\bibinfo{year}{2009}),
  \urlprefix\url{http://link.aps.org/doi/10.1103/PhysRevB.79.195440}.

\bibitem[{\citenamefont{Li et~al.}(2015)\citenamefont{Li, Hudson, Dzurak, and
  Hamilton}}]{Li2015}
\bibinfo{author}{\bibfnamefont{R.}~\bibnamefont{Li}},
  \bibinfo{author}{\bibfnamefont{F.~E.} \bibnamefont{Hudson}},
  \bibinfo{author}{\bibfnamefont{A.~S.} \bibnamefont{Dzurak}},
  \bibnamefont{and} \bibinfo{author}{\bibfnamefont{A.~R.}
  \bibnamefont{Hamilton}}, \bibinfo{journal}{Nano Lett.}
  \textbf{\bibinfo{volume}{15}}, \bibinfo{pages}{7314} (\bibinfo{year}{2015}),
  ISSN \bibinfo{issn}{1530-6984},
  \urlprefix\url{http://dx.doi.org/10.1021/acs.nanolett.5b02561}.

\bibitem[{\citenamefont{Voisin et~al.}(2016)\citenamefont{Voisin, Maurand,
  Barraud, Vinet, Jehl, Sanquer, Renard, and De~Franceschi}}]{Voisin2016}
\bibinfo{author}{\bibfnamefont{B.}~\bibnamefont{Voisin}},
  \bibinfo{author}{\bibfnamefont{R.}~\bibnamefont{Maurand}},
  \bibinfo{author}{\bibfnamefont{S.}~\bibnamefont{Barraud}},
  \bibinfo{author}{\bibfnamefont{M.}~\bibnamefont{Vinet}},
  \bibinfo{author}{\bibfnamefont{X.}~\bibnamefont{Jehl}},
  \bibinfo{author}{\bibfnamefont{M.}~\bibnamefont{Sanquer}},
  \bibinfo{author}{\bibfnamefont{J.}~\bibnamefont{Renard}}, \bibnamefont{and}
  \bibinfo{author}{\bibfnamefont{S.}~\bibnamefont{De~Franceschi}},
  \bibinfo{journal}{Nano Lett.} \textbf{\bibinfo{volume}{16}},
  \bibinfo{pages}{88} (\bibinfo{year}{2016}), ISSN \bibinfo{issn}{1530-6984},
  \urlprefix\url{http://dx.doi.org/10.1021/acs.nanolett.5b02920}.

\bibitem[{\citenamefont{Maurand et~al.}(2016)\citenamefont{Maurand, Jehl,
  Kotekar~Patil, Corna, Bohuslavskyi, Lavieville, Hutin, Barraud, Vinet,
  Sanquer et~al.}}]{Maurand2016}
\bibinfo{author}{\bibfnamefont{R.}~\bibnamefont{Maurand}},
  \bibinfo{author}{\bibfnamefont{X.}~\bibnamefont{Jehl}},
  \bibinfo{author}{\bibfnamefont{D.}~\bibnamefont{Kotekar~Patil}},
  \bibinfo{author}{\bibfnamefont{A.}~\bibnamefont{Corna}},
  \bibinfo{author}{\bibfnamefont{H.}~\bibnamefont{Bohuslavskyi}},
  \bibinfo{author}{\bibfnamefont{R.}~\bibnamefont{Lavieville}},
  \bibinfo{author}{\bibfnamefont{L.}~\bibnamefont{Hutin}},
  \bibinfo{author}{\bibfnamefont{S.}~\bibnamefont{Barraud}},
  \bibinfo{author}{\bibfnamefont{M.}~\bibnamefont{Vinet}},
  \bibinfo{author}{\bibfnamefont{M.}~\bibnamefont{Sanquer}},
  \bibnamefont{et~al.}, \bibinfo{journal}{arXiv:1605.07599}
  (\bibinfo{year}{2016}).

\bibitem[{\citenamefont{Barraud et~al.}(2012)\citenamefont{Barraud, Coquand,
  Casse, Koyama, Hartmann, Maffini-Alvaro, Comboroure, Vizioz, Aussenac, Faynot
  et~al.}}]{Barraud2012a}
\bibinfo{author}{\bibfnamefont{S.}~\bibnamefont{Barraud}},
  \bibinfo{author}{\bibfnamefont{R.}~\bibnamefont{Coquand}},
  \bibinfo{author}{\bibfnamefont{M.}~\bibnamefont{Casse}},
  \bibinfo{author}{\bibfnamefont{M.}~\bibnamefont{Koyama}},
  \bibinfo{author}{\bibfnamefont{J.}~\bibnamefont{Hartmann}},
  \bibinfo{author}{\bibfnamefont{V.}~\bibnamefont{Maffini-Alvaro}},
  \bibinfo{author}{\bibfnamefont{C.}~\bibnamefont{Comboroure}},
  \bibinfo{author}{\bibfnamefont{C.}~\bibnamefont{Vizioz}},
  \bibinfo{author}{\bibfnamefont{F.}~\bibnamefont{Aussenac}},
  \bibinfo{author}{\bibfnamefont{O.}~\bibnamefont{Faynot}},
  \bibnamefont{et~al.}, \bibinfo{journal}{Electron Device Letters, IEEE}
  \textbf{\bibinfo{volume}{33}}, \bibinfo{pages}{1526} (\bibinfo{year}{2012}),
  ISSN \bibinfo{issn}{0741-3106}.

\bibitem[{\citenamefont{Barraud et~al.}(2016)\citenamefont{Barraud, Lavieville,
  Hutin, Bohuslavskyi, Vinet, Corna, Clapera, Sanquer, and Jehl}}]{Barraud2016}
\bibinfo{author}{\bibfnamefont{S.}~\bibnamefont{Barraud}},
  \bibinfo{author}{\bibfnamefont{R.}~\bibnamefont{Lavieville}},
  \bibinfo{author}{\bibfnamefont{L.}~\bibnamefont{Hutin}},
  \bibinfo{author}{\bibfnamefont{H.}~\bibnamefont{Bohuslavskyi}},
  \bibinfo{author}{\bibfnamefont{M.}~\bibnamefont{Vinet}},
  \bibinfo{author}{\bibfnamefont{A.}~\bibnamefont{Corna}},
  \bibinfo{author}{\bibfnamefont{P.}~\bibnamefont{Clapera}},
  \bibinfo{author}{\bibfnamefont{M.}~\bibnamefont{Sanquer}}, \bibnamefont{and}
  \bibinfo{author}{\bibfnamefont{X.}~\bibnamefont{Jehl}},
  \bibinfo{journal}{Technologies} \textbf{\bibinfo{volume}{4}},
  \bibinfo{pages}{10} (\bibinfo{year}{2016}), ISSN \bibinfo{issn}{2227-7080},
  \urlprefix\url{http://www.mdpi.com/2227-7080/4/1/10}.

\bibitem[{\citenamefont{Danon and Nazarov}(2009)}]{Danon2009}
\bibinfo{author}{\bibfnamefont{J.}~\bibnamefont{Danon}} \bibnamefont{and}
  \bibinfo{author}{\bibfnamefont{Y.~V.} \bibnamefont{Nazarov}},
  \bibinfo{journal}{Phys. Rev. B} \textbf{\bibinfo{volume}{80}},
  \bibinfo{pages}{041301} (\bibinfo{year}{2009}),
  \urlprefix\url{http://link.aps.org/doi/10.1103/PhysRevB.80.041301}.

\bibitem[{\citenamefont{Nadj-Perge
  et~al.}(2010{\natexlab{b}})\citenamefont{Nadj-Perge, Frolov, van Tilburg,
  Danon, Nazarov, Algra, Bakkers, and Kouwenhoven}}]{Nadj-Perge2010}
\bibinfo{author}{\bibfnamefont{S.}~\bibnamefont{Nadj-Perge}},
  \bibinfo{author}{\bibfnamefont{S.~M.} \bibnamefont{Frolov}},
  \bibinfo{author}{\bibfnamefont{J.~W.~W.} \bibnamefont{van Tilburg}},
  \bibinfo{author}{\bibfnamefont{J.}~\bibnamefont{Danon}},
  \bibinfo{author}{\bibfnamefont{Y.~V.} \bibnamefont{Nazarov}},
  \bibinfo{author}{\bibfnamefont{R.}~\bibnamefont{Algra}},
  \bibinfo{author}{\bibfnamefont{E.~P. A.~M.} \bibnamefont{Bakkers}},
  \bibnamefont{and} \bibinfo{author}{\bibfnamefont{L.~P.}
  \bibnamefont{Kouwenhoven}}, \bibinfo{journal}{Phys. Rev. B}
  \textbf{\bibinfo{volume}{81}}, \bibinfo{pages}{201305}
  (\bibinfo{year}{2010}{\natexlab{b}}),
  \urlprefix\url{http://link.aps.org/doi/10.1103/PhysRevB.81.201305}.

\bibitem[{\citenamefont{Kotekar-Patil et~al.}()\citenamefont{Kotekar-Patil,
  Corna, Maurand, Crippa, Orlov, Barraud, Jehl, De~Franceschi, and
  Sanquer}}]{Kotekar2016}
\bibinfo{author}{\bibfnamefont{D.}~\bibnamefont{Kotekar-Patil}},
  \bibinfo{author}{\bibfnamefont{A.}~\bibnamefont{Corna}},
  \bibinfo{author}{\bibfnamefont{R.}~\bibnamefont{Maurand}},
  \bibinfo{author}{\bibfnamefont{A.}~\bibnamefont{Crippa}},
  \bibinfo{author}{\bibfnamefont{A.}~\bibnamefont{Orlov}},
  \bibinfo{author}{\bibfnamefont{S.}~\bibnamefont{Barraud}},
  \bibinfo{author}{\bibfnamefont{X.}~\bibnamefont{Jehl}},
  \bibinfo{author}{\bibfnamefont{S.}~\bibnamefont{De~Franceschi}},
  \bibnamefont{and} \bibinfo{author}{\bibfnamefont{M.}~\bibnamefont{Sanquer}},
  \bibinfo{journal}{arXiv:1606.05855v1}  (????).

\bibitem[{\citenamefont{van~der Wiel et~al.}(2002)\citenamefont{van~der Wiel,
  De~Franceschi, Elzerman, Fujisawa, Tarucha, and Kouwenhoven}}]{Wiel2002}
\bibinfo{author}{\bibfnamefont{W.~G.} \bibnamefont{van~der Wiel}},
  \bibinfo{author}{\bibfnamefont{S.}~\bibnamefont{De~Franceschi}},
  \bibinfo{author}{\bibfnamefont{J.~M.} \bibnamefont{Elzerman}},
  \bibinfo{author}{\bibfnamefont{T.}~\bibnamefont{Fujisawa}},
  \bibinfo{author}{\bibfnamefont{S.}~\bibnamefont{Tarucha}}, \bibnamefont{and}
  \bibinfo{author}{\bibfnamefont{L.~P.} \bibnamefont{Kouwenhoven}},
  \bibinfo{journal}{Rev. Mod. Phys.} \textbf{\bibinfo{volume}{75}},
  \bibinfo{pages}{1} (\bibinfo{year}{2002}),
  \urlprefix\url{http://link.aps.org/doi/10.1103/RevModPhys.75.1}.

\bibitem[{\citenamefont{Golovach et~al.}(2011)\citenamefont{Golovach, Jehl,
  Houzet, Pierre, Roche, Sanquer, and Glazman}}]{Golovach2011}
\bibinfo{author}{\bibfnamefont{V.~N.} \bibnamefont{Golovach}},
  \bibinfo{author}{\bibfnamefont{X.}~\bibnamefont{Jehl}},
  \bibinfo{author}{\bibfnamefont{M.}~\bibnamefont{Houzet}},
  \bibinfo{author}{\bibfnamefont{M.}~\bibnamefont{Pierre}},
  \bibinfo{author}{\bibfnamefont{B.}~\bibnamefont{Roche}},
  \bibinfo{author}{\bibfnamefont{M.}~\bibnamefont{Sanquer}}, \bibnamefont{and}
  \bibinfo{author}{\bibfnamefont{L.~I.} \bibnamefont{Glazman}},
  \bibinfo{journal}{Phys. Rev. B} \textbf{\bibinfo{volume}{83}},
  \bibinfo{pages}{075401} (\bibinfo{year}{2011}),
  \urlprefix\url{http://link.aps.org/doi/10.1103/PhysRevB.83.075401}.

\bibitem[{\citenamefont{Niquet et~al.}(2009)\citenamefont{Niquet, Rideau,
  Tavernier, Jaouen, and Blase}}]{Niquet2009}
\bibinfo{author}{\bibfnamefont{Y.~M.} \bibnamefont{Niquet}},
  \bibinfo{author}{\bibfnamefont{D.}~\bibnamefont{Rideau}},
  \bibinfo{author}{\bibfnamefont{C.}~\bibnamefont{Tavernier}},
  \bibinfo{author}{\bibfnamefont{H.}~\bibnamefont{Jaouen}}, \bibnamefont{and}
  \bibinfo{author}{\bibfnamefont{X.}~\bibnamefont{Blase}},
  \bibinfo{journal}{Phys. Rev. B} \textbf{\bibinfo{volume}{79}},
  \bibinfo{pages}{245201} (\bibinfo{year}{2009}),
  \urlprefix\url{http://link.aps.org/doi/10.1103/PhysRevB.79.245201}.

\bibitem[{\citenamefont{Voisin et~al.}(2014)\citenamefont{Voisin, Nguyen,
  Renard, Jehl, Barraud, Triozon, Vinet, Duchemin, Niquet, de~Franceschi
  et~al.}}]{edges_NL2014}
\bibinfo{author}{\bibfnamefont{B.}~\bibnamefont{Voisin}},
  \bibinfo{author}{\bibfnamefont{V.-H.} \bibnamefont{Nguyen}},
  \bibinfo{author}{\bibfnamefont{J.}~\bibnamefont{Renard}},
  \bibinfo{author}{\bibfnamefont{X.}~\bibnamefont{Jehl}},
  \bibinfo{author}{\bibfnamefont{S.}~\bibnamefont{Barraud}},
  \bibinfo{author}{\bibfnamefont{F.}~\bibnamefont{Triozon}},
  \bibinfo{author}{\bibfnamefont{M.}~\bibnamefont{Vinet}},
  \bibinfo{author}{\bibfnamefont{I.}~\bibnamefont{Duchemin}},
  \bibinfo{author}{\bibfnamefont{Y.-M.} \bibnamefont{Niquet}},
  \bibinfo{author}{\bibfnamefont{S.}~\bibnamefont{de~Franceschi}},
  \bibnamefont{et~al.}, \bibinfo{journal}{Nano letters}
  \textbf{\bibinfo{volume}{14}}, \bibinfo{pages}{2094} (\bibinfo{year}{2014}).

\bibitem[{\citenamefont{Pierre et~al.}(2009)\citenamefont{Pierre, Hofheinz,
  Jehl, Sanquer, Molas, Vinet, and Deleonibus}}]{Pierre2009}
\bibinfo{author}{\bibfnamefont{M.}~\bibnamefont{Pierre}},
  \bibinfo{author}{\bibfnamefont{M.}~\bibnamefont{Hofheinz}},
  \bibinfo{author}{\bibfnamefont{X.}~\bibnamefont{Jehl}},
  \bibinfo{author}{\bibfnamefont{M.}~\bibnamefont{Sanquer}},
  \bibinfo{author}{\bibfnamefont{G.}~\bibnamefont{Molas}},
  \bibinfo{author}{\bibfnamefont{M.}~\bibnamefont{Vinet}}, \bibnamefont{and}
  \bibinfo{author}{\bibfnamefont{S.}~\bibnamefont{Deleonibus}},
  \bibinfo{journal}{Eur. Phys. J. B} \textbf{\bibinfo{volume}{70}},
  \bibinfo{pages}{475} (\bibinfo{year}{2009}),
  \urlprefix\url{http://dx.doi.org/10.1140/epjb/e2009-00258-4}.

\bibitem[{\citenamefont{Hofheinz et~al.}(2006)\citenamefont{Hofheinz, Jehl,
  Sanquer, Molas, Vinet, and Deleonibus}}]{Hofheinz2006}
\bibinfo{author}{\bibfnamefont{M.}~\bibnamefont{Hofheinz}},
  \bibinfo{author}{\bibfnamefont{X.}~\bibnamefont{Jehl}},
  \bibinfo{author}{\bibfnamefont{M.}~\bibnamefont{Sanquer}},
  \bibinfo{author}{\bibfnamefont{G.}~\bibnamefont{Molas}},
  \bibinfo{author}{\bibfnamefont{M.}~\bibnamefont{Vinet}}, \bibnamefont{and}
  \bibinfo{author}{\bibfnamefont{S.}~\bibnamefont{Deleonibus}},
  \bibinfo{journal}{Applied Physics Letters} \textbf{\bibinfo{volume}{89}},
  \bibinfo{pages}{143504} (\bibinfo{year}{2006}).

\bibitem[{\citenamefont{Ono et~al.}(2002)\citenamefont{Ono, Austing, Tokura,
  and Tarucha}}]{Ono2002}
\bibinfo{author}{\bibfnamefont{K.}~\bibnamefont{Ono}},
  \bibinfo{author}{\bibfnamefont{D.~G.} \bibnamefont{Austing}},
  \bibinfo{author}{\bibfnamefont{Y.}~\bibnamefont{Tokura}}, \bibnamefont{and}
  \bibinfo{author}{\bibfnamefont{S.}~\bibnamefont{Tarucha}},
  \bibinfo{journal}{Science} \textbf{\bibinfo{volume}{297}},
  \bibinfo{pages}{1313} (\bibinfo{year}{2002}),
  \urlprefix\url{http://science.sciencemag.org/content/297/5585/1313.abstract}.

\bibitem[{\citenamefont{Yamahata et~al.}(2012)\citenamefont{Yamahata, Kodera,
  Churchill, Uchida, Marcus, and Oda}}]{Yamahata2012}
\bibinfo{author}{\bibfnamefont{G.}~\bibnamefont{Yamahata}},
  \bibinfo{author}{\bibfnamefont{T.}~\bibnamefont{Kodera}},
  \bibinfo{author}{\bibfnamefont{H.~O.~H.} \bibnamefont{Churchill}},
  \bibinfo{author}{\bibfnamefont{K.}~\bibnamefont{Uchida}},
  \bibinfo{author}{\bibfnamefont{C.~M.} \bibnamefont{Marcus}},
  \bibnamefont{and} \bibinfo{author}{\bibfnamefont{S.}~\bibnamefont{Oda}},
  \bibinfo{journal}{Phys. Rev. B} \textbf{\bibinfo{volume}{86}},
  \bibinfo{pages}{115322} (\bibinfo{year}{2012}),
  \urlprefix\url{http://link.aps.org/doi/10.1103/PhysRevB.86.115322}.

\bibitem[{\citenamefont{Lai et~al.}(2011)\citenamefont{Lai, Lim, Yang,
  Zwanenburg, Coish, Qassemi, Morello, and Dzurak}}]{Morello_2011SB}
\bibinfo{author}{\bibfnamefont{N.}~\bibnamefont{Lai}},
  \bibinfo{author}{\bibfnamefont{W.}~\bibnamefont{Lim}},
  \bibinfo{author}{\bibfnamefont{C.}~\bibnamefont{Yang}},
  \bibinfo{author}{\bibfnamefont{F.}~\bibnamefont{Zwanenburg}},
  \bibinfo{author}{\bibfnamefont{W.}~\bibnamefont{Coish}},
  \bibinfo{author}{\bibfnamefont{F.}~\bibnamefont{Qassemi}},
  \bibinfo{author}{\bibfnamefont{A.}~\bibnamefont{Morello}}, \bibnamefont{and}
  \bibinfo{author}{\bibfnamefont{A.}~\bibnamefont{Dzurak}},
  \bibinfo{journal}{Scientific reports} \textbf{\bibinfo{volume}{1}}
  (\bibinfo{year}{2011}).

\bibitem[{\citenamefont{Pfund et~al.}(2007)\citenamefont{Pfund, Shorubalko,
  Ensslin, and Leturcq}}]{Pfund2007}
\bibinfo{author}{\bibfnamefont{A.}~\bibnamefont{Pfund}},
  \bibinfo{author}{\bibfnamefont{I.}~\bibnamefont{Shorubalko}},
  \bibinfo{author}{\bibfnamefont{K.}~\bibnamefont{Ensslin}}, \bibnamefont{and}
  \bibinfo{author}{\bibfnamefont{R.}~\bibnamefont{Leturcq}},
  \bibinfo{journal}{Phys. Rev. Lett.} \textbf{\bibinfo{volume}{99}},
  \bibinfo{pages}{036801} (\bibinfo{year}{2007}),
  \urlprefix\url{http://link.aps.org/doi/10.1103/PhysRevLett.99.036801}.

\bibitem[{\citenamefont{Pribiag et~al.}(2013)\citenamefont{Pribiag, Nadj-Perge,
  Frolov, van~den Berg, van Weperen, Plissard, Bakkers, and
  Kouwenhoven}}]{S.2013}
\bibinfo{author}{\bibfnamefont{S.~V.} \bibnamefont{Pribiag}},
  \bibinfo{author}{\bibfnamefont{S.}~\bibnamefont{Nadj-Perge}},
  \bibinfo{author}{\bibfnamefont{S.~M.} \bibnamefont{Frolov}},
  \bibinfo{author}{\bibfnamefont{J.~W.~G.} \bibnamefont{van~den Berg}},
  \bibinfo{author}{\bibfnamefont{I.}~\bibnamefont{van Weperen}},
  \bibinfo{author}{\bibfnamefont{S.~R.} \bibnamefont{Plissard}},
  \bibinfo{author}{\bibfnamefont{E.~P. A.~M.} \bibnamefont{Bakkers}},
  \bibnamefont{and} \bibinfo{author}{\bibfnamefont{L.~P.}
  \bibnamefont{Kouwenhoven}}, \bibinfo{journal}{Nat Nano}
  \textbf{\bibinfo{volume}{8}}, \bibinfo{pages}{170} (\bibinfo{year}{2013}),
  ISSN \bibinfo{issn}{1748-3387},
  \urlprefix\url{http://dx.doi.org/10.1038/nnano.2013.5}.

\bibitem[{\citenamefont{Nadj-Perge et~al.}(2012)\citenamefont{Nadj-Perge,
  Pribiag, van~den Berg, Zuo, Plissard, Bakkers, Frolov, and
  Kouwenhoven}}]{Nadj-Perge2012}
\bibinfo{author}{\bibfnamefont{S.}~\bibnamefont{Nadj-Perge}},
  \bibinfo{author}{\bibfnamefont{V.~S.} \bibnamefont{Pribiag}},
  \bibinfo{author}{\bibfnamefont{J.~W.~G.} \bibnamefont{van~den Berg}},
  \bibinfo{author}{\bibfnamefont{K.}~\bibnamefont{Zuo}},
  \bibinfo{author}{\bibfnamefont{S.~R.} \bibnamefont{Plissard}},
  \bibinfo{author}{\bibfnamefont{E.~P. A.~M.} \bibnamefont{Bakkers}},
  \bibinfo{author}{\bibfnamefont{S.~M.} \bibnamefont{Frolov}},
  \bibnamefont{and} \bibinfo{author}{\bibfnamefont{L.~P.}
  \bibnamefont{Kouwenhoven}}, \bibinfo{journal}{Phys. Rev. Lett.}
  \textbf{\bibinfo{volume}{108}}, \bibinfo{pages}{166801}
  (\bibinfo{year}{2012}),
  \urlprefix\url{http://link.aps.org/doi/10.1103/PhysRevLett.108.166801}.

\end{thebibliography}

\end{document}